\documentclass[a4paper,11pt]{article}

\usepackage{jinstpub} 

\usepackage[utf8]{inputenc}
\usepackage{siunitx}
\usepackage{multirow}
\usepackage{graphicx}
\usepackage{mathrsfs}
\graphicspath{ {./figures/} }
\usepackage{caption}
\usepackage{subcaption}
\usepackage{color}
\usepackage[acronym]{glossaries}

\usepackage{lineno}

\title{\boldmath Neural-network-based level-1 trigger upgrade for the SuperCDMS experiment at SNOLAB}

\author[a,1]{H.~Meyer~zu~Theenhausen,\note{Corresponding author.}}
\author[a]{B.~von~Krosigk,}
\author[b]{J.~S.~Wilson}

\affiliation[a]{Institute for Astroparticle Physics (IAP), Karlsruhe Institute of Technology (KIT),\\76344 Eggenstein-Leopoldshafen, Germany}
\affiliation[b]{Baylor University,\\Waco 76706, TX, USA}

\emailAdd{hanno.theenhausen@kit.edu}

\abstract{
    The extended physics program of the SuperCDMS SNOLAB dark matter search experiment aims to maximize the sensitivity to low-mass dark matter. To realize this, an upgrade of the existing level-1 trigger of the data acquisition system is proposed by making use of a recurrent neural network to be implemented on the trigger FPGA. This provides an improved amplitude estimator and signal-noise discriminator based on the combined information of filtered traces from individual detector channels. The architecture and configuration of this neural trigger are discussed in this article, and the improvements in key performance indicators such as the efficiency, resolution, and noise rate are quantified based on signal simulations and noise data. Based on the findings in this proof of concept, the trigger threshold is expected to be lowered by ${\sim}22$\%.
}


\makeglossaries
\begin{document}
\maketitle
\flushbottom

\newacronym{HV}{HV}{High Voltage (detector)}
\newacronym{iZIP}{iZIP}{interleaved Z-sensitive Ionization Phonon (detector)}
\newacronym{DCRC}{DCRC}{Detector Control and Readout Card}
\newacronym{FPGA}{FPGA}{Field-Programmable Gate Array}
\newacronym{FIR}{FIR}{Finite-Impulse-Response (filter)}
\newacronym{OF}{OF}{Optimal Filter}
\newacronym{PT}{PT}{Phonon Total (channel)}
\newacronym{LSTM}{LSTM}{Long Short-Term Memory}
\newacronym{NN}{NN}{Neural Network}
\newacronym{ADC}{ADC}{Analog-to-Digital Converter}
\newacronym{FOF}{FOF}{Flattened Optimal Filter}
\newacronym{DMC}{DMC}{Detector Monte Carlo}
\newacronym{WIMP}{WIMP}{Weakly Interacting Massive Particle}
\newacronym{PSD}{PSD}{Power Spectral Density}
\newacronym{FIFO}{FIFO}{First-In-First-Out (buffer)}

\printglossary[type=\acronymtype,style=index,nonumberlist,nogroupskip]

\section{Introduction}
\label{sec:intro}
SuperCDMS SNOLAB is a cryogenic dark matter search experiment based on silicon and germanium crystal detectors~\cite{Agnese_2017}. Potential interactions of dark matter particles with the detector target material produce phonon and charge signals which are detected via transition edge sensors and charge electrodes, respectively. Two types of detectors are employed: the interleaved Z-sensitive Ionization Phonon detector (\acrshort{iZIP}), which uses 12 phonon and 4 charge channels, and the high voltage (\acrshort{HV}) detector, which omits the charge channels. The \acrshort{iZIP} detectors allow for discrimination between nuclear and electron recoils by observing the ratio of the phonon and ionization yields. The \acrshort{HV} detectors apply a high bias voltage, resulting in large Neganov-Trofimov-Luke amplification~\cite{Neganov:1985khw,Luke1988VoltageassistedCI} and improved sensitivity of the phonon signal. The channels from an \acrshort{iZIP} detector are connected to a Detector Control and Readout Card (\acrshort{DCRC}) which contains the digitizers and the level-1 (L1) trigger system~\cite{Wilson_2022} on a Field-Programmable Gate Array (\acrshort{FPGA}). An \acrshort{HV} detector is attached to two \acrshort{DCRC}s, each connected to the 6 phonon channels from one detector side. From the channels the L1 trigger extracts signal pulses from the noise baseline which contains components directly originated from electronic noise sources, and components that have been converted from vibrational sources. Reaching the highest possible trigger efficiency at the lowest possible threshold while maintaining an acceptable noise rate is crucial to extend the science goals of SuperCDMS and other rare event searches~\cite{MANCUSO2019492}. At the heart of the SuperCDMS L1 trigger is a finite-impulse-response (\acrshort{FIR}) filter configured as an optimal filter (\acrshort{OF}), a minimum-variance estimator of the amplitude of a pulse with the shape modeled by a template in the presence of stationary noise~\cite{OF_gatti}. In the default configuration of the L1 trigger system, the \acrshort{OF} \acrshort{FIR} is applied to the phonon total (\acrshort{PT}) channel, i.e. the sum of all 12 (6) phonon channels of an \acrshort{iZIP} detector (\acrshort{HV} detector side). For a germanium \acrshort{iZIP} (\acrshort{HV}) detector this results in an efficient trigger for pulses with a nuclear recoil energy of $\sim$272 (40) eV at a negligible noise rate~\cite{Agnese_2017}.

In this article an upgrade to the L1 trigger system toward a ``neural trigger'' is proposed, making use of a Long Short-Term Memory \cite{hochreiter1997long} (\acrshort{LSTM}) neural network (\acrshort{NN}) that will be implemented on the trigger \acrshort{FPGA}.
The inputs to this \acrshort{NN} are \acrshort{OF} filtered traces from the \acrshort{PT} channel, as well as \acrshort{OF} filtered traces from individual phonon channels using templates describing variations in pulse shapes. The inputs are combined in the \acrshort{NN}, which is trained to output a high-resolution amplitude estimation. The filtered input traces are also preserved to enable the legacy operation of the trigger.
Compared to the default \acrshort{PT} channel \acrshort{OF}, the neural trigger has two advantages which offer the potential to significantly improve the signal-noise discrimination. 
First, the \acrshort{NN} can learn physical combinations of different fast and slow rising signal pulse shape components of each channel, depending on the position of the particle interaction within the detector and the proximity to the respective detector channels. The physical, i.e. signal-like combinations can then be distinguished from non-physical, i.e. noise-like combinations which improves the signal-noise discrimination. To give an example, the presence of a dominant fast rising pulse component in only one channel and slow rising pulse components in the remaining channels represents a signal-like topology that the \acrshort{NN} should be able to distinguish from noise-like topologies. Secondly, using the channel-specific inputs, the \acrshort{NN} can learn to exploit information about noise correlations among the channels. 
Simulated pulses and noise data are used to train the \acrshort{NN} and to evaluate the performance. A baseline inversion method is employed to mitigate false pulse labeling from residual pulses in the noise data. With this trigger upgrade, the threshold can be lowered by 22\% without degrading the efficiency or the noise rejection. The data samples used allow for a good approximation of the real detector conditions, however more studies are needed to validate the results on a broader basis. The neural trigger architecture and configuration for an \acrshort{iZIP} detector are detailed in section~\ref{sec:neuraltrigger}. In section~\ref{sec:inputdata}, based on noise traces from a germanium \acrshort{iZIP} detector and simulated pulses, the full data pipeline through the trigger modules is illustrated from the input data to the \acrshort{FIR} filter and through the \acrshort{NN}. The performance of the upgraded neural trigger in this scenario is quantified and compared to the default trigger in section~\ref{sec:performance}. Finally, section~\ref{sec:conclusion} gives a conclusion of the results.

\section{Neural Trigger}
\label{sec:neuraltrigger}
In the proposed neural trigger upgrade for the SuperCDMS L1 trigger, a new \acrshort{NN} module is inserted in the modular structure of the default trigger~\cite{Wilson_2022} within the \acrshort{FPGA} via high-level synthesis. Additionally, the number of trigger paths (parallel trigger modules serving as inputs to the \acrshort{NN} module) is considerably increased. Each trigger path is configured with a specific channel input and \acrshort{FIR} filter type. The proposed neural trigger architecture and configuration for an \acrshort{iZIP} detector are outlined in this section.

\subsection{Architecture}

The default L1 trigger forms a data pipeline consisting of several discrete modules in parallel trigger paths. In each trigger path in the downsampling filter, linear combination, and \acrshort{FIR} modules, the trigger input data from the Analog-to-Digital Converters (\acrshort{ADC}s) of every channel are downsampled, linearly combined and filtered. In the subsequent threshold logic and peak search modules, the filtered traces are subjected to a configurable threshold logic and, for each threshold crossing, a ``trigger window'' is defined, during which a peak search algorithm records the peak amplitude, peak time, and threshold information as a set of trigger primitives. The final trigger decisions are made in the trigger logic module based on Boolean combinations of the trigger primitives. An extensive description of the existing L1 trigger modules is detailed in ref.~\cite{Wilson_2022}.

\begin{figure}[b]
    \centering
    \includegraphics[width=1.\textwidth]{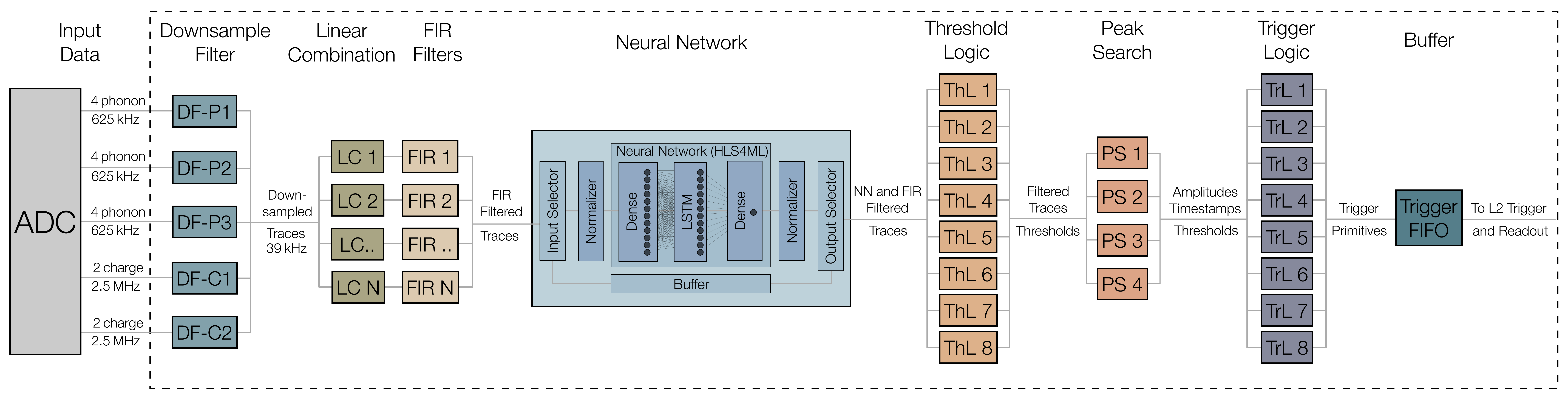}
    \caption{Layout of the SuperCDMS level-1 trigger upgrade proposal with emphasis on the neural network module. The number of parallel trigger paths $N$ is set to 26. The neural network contains dense and long short-term memory (LSTM) neurons and will be implemented in the FPGA firmware via high-level synthesis for machine learning (HLS4ML).}
    \label{fig:neuraltriggerlayout}
\end{figure}

In the neural trigger upgrade the \acrshort{NN} module is placed between the \acrshort{FIR} filter module and the threshold logic module. Figure~\ref{fig:neuraltriggerlayout} shows the \acrshort{NN} module within the existing trigger layout and illustrates its internal structure schematically. The number of trigger paths $N$ is increased from originally $N=4$ to $N=26$. The \acrshort{NN} module receives the filtered traces from the \acrshort{FIR} module for each trigger path as its inputs. Within the \acrshort{NN} module a configurable 26-bit register value determines which of the 26 trigger paths will be input to the network. The chosen inputs are normalized to the order of 1 via a bitshift operation. The \acrshort{NN} consists of one dense layer with 24 neurons, one \acrshort{LSTM} layer with 12 neurons, and 1 dense output layer. The first dense layer and the \acrshort{LSTM} layer use the ReLU \cite{agarap2018deep} output activation function and the dense output layer uses a linear activation function. The weights in each layer, represented by 32 bits of which 16 bits each represent the signed number above and below the binary point, are determined in an offline training procedure detailed in section~\ref{sec:training}. The \acrshort{NN} output is normalized back to its original scale. During the computation within the \acrshort{NN}, the \acrshort{NN} module input data are stored in a First-In-First-Out (\acrshort{FIFO}) buffer which is read out once the \acrshort{NN} produces an output and then is transferred to the \acrshort{NN} module output. The \acrshort{NN} output overwrites the data in a desired trigger path that can be configured via a 6-bit output selector register value. The output of the \acrshort{NN} module is sent further to the threshold logic module. A software implementation of the \acrshort{NN} module has been developed in python for the trigger simulation which is used for the studies in this article. Therein the \acrshort{NN} layers were compiled in a sequential model using the Keras API~\cite{chollet2015keras} and converted to an HLS4ML ~\cite{Duarte_2018} model allowing for the emulation of the implementation of the \acrshort{NN} on the \acrshort{FPGA} to bitwise precision. The specific \acrshort{NN} architecture was chosen because two layers with the ReLU activation function allow the approximation of arbitrary non-linear functions to a good degree~\cite{165599} and the \acrshort{LSTM} units are able to account for time correlations. The numbers and types of nodes are motivated by the number of pulse shape components per single channel and a coarse optimization study. The \acrshort{NN} architecture is tentative and may slightly change in the course of future studies. Upcoming HLS4ML releases may support QLSTM layers provided by QKeras~\cite{https://doi.org/10.48550/arxiv.2006.10159}, which should provide quantization-aware training and thereby a further improvement of this implementation.

During the design of the readout electronics a possible trigger upgrade has already been foreseen and an \acrshort{FPGA} with significant spare resources was employed. The 100\,MHz \acrshort{FPGA} clock is significantly faster than the 39\,kHz sample rate, so time-division multiplexing is employed to keep the resource usage small. Using the default trigger design the \acrshort{FPGA} utilizes around 33/18/14\% of the logic-elements/memory/multiplier resources and with the neural trigger the \acrshort{FPGA} is estimated to utilize around 72/76/95\% of these.

\subsection{Configuration}

In the proposed configuration of the neural trigger, the first two trigger paths use the \acrshort{PT} channel. This requires the linear combination coefficients in each of the 12 phonon channels to be set to the same value, which is chosen to be the maximum 8-bit value, 127. This assumes a similar gain in all channels. In case of varying gains the linear combination coefficients can also differ. The \acrshort{FIR} coefficients in the first trigger path resemble a time-domain \acrshort{OF}.
As was shown in ref.~\cite{Wilson_2022}, oscillations in the \acrshort{OF} \acrshort{FIR} coefficient sidebands can lead to artifact echo triggers. To suppress these echos, the \acrshort{FIR} filter coefficients in the second trigger path are a ``flattened OF'' (\acrshort{FOF}) where the oscillations in the sidebands have been flattened. The remaining 24 trigger paths alternate which single phonon-channel coefficient is set to 127 while all other coefficients are kept at zero. This results in two trigger paths per single phonon channel - each configured with \acrshort{OF} \acrshort{FIR} coefficients based on the channel-specific fast and slow pulse shape components, respectively. All \acrshort{FIR} outputs are configured to enter the \acrshort{NN} and the \acrshort{NN} output is configured to be written to an available trigger path, for which the third trigger path is chosen. The output of the first two trigger paths are part of the legacy information as a backup option. The nominal thresholds in the threshold logic module are applied to the third trigger path containing the \acrshort{NN} output. 
Table~\ref{tab:config} summarizes the configuration in each trigger path for an \acrshort{iZIP} detector. In case of an \acrshort{HV} detector, only 6 phonon channels are available and the number of used trigger paths is reduced accordingly.

\begin{table}[h]
\center
\begin{tabular}{|c|c|c|}
\hline
Trigger Path          & Detector Channels           & FIR Filter                    \\ \hline
1                     & Phonon Total         & Optimal Filter                \\ 
2                     & Phonon Total         & Flattened Optimal Filter      \\ 
3, .., 14             & Single Phonon  1, .., 12 & Fast Component Optimal Filter \\ 
15, .., 26            & Single Phonon  1, .., 12 & Slow Component Optimal Filter \\ \hline

\end{tabular}
\caption{Summary of the proposed linear combination and FIR filter configuration for the neural trigger in each trigger path.}
\label{tab:config}
\end{table}

\section{Data Pipeline}
\label{sec:inputdata}
The neural trigger forms a data pipeline consisting of discrete modules, analogous to the default L1 trigger, but with an additional \acrshort{NN} module inserted.
The functioning of the neural trigger in its proposed configuration is demonstrated using trigger simulations at various stages of the data pipeline. The data are propagated stepwise only up to certain stages of the pipeline in order to derive the correct settings for the stages after. In the first step the data at the \acrshort{ADC} level are propagated through the downsampling and linear combination modules to derive the \acrshort{OF} coefficients for the \acrshort{FIR} module. In the next step the data are propagated through the downsampling, linear combination and \acrshort{FIR} modules in order to produce the inputs necessary to train a model for the \acrshort{NN} module. After the model has been loaded to the \acrshort{NN} module, the data can be propagated through the full data pipeline and the performance of the neural trigger can be evaluated. In this section this stepwise procedure is demonstrated using input data from simulated pulses and noise traces measured with a prototype detector.

\subsection{Input Data}

The input data in this evaluation consist of SuperCDMS Detector Monte Carlo (\acrshort{DMC}) phonon pulses injected into randomly triggered noise traces measured with an \acrshort{iZIP} detector.

\begin{figure}[t]
     \centering
     \begin{subfigure}[b]{0.49\textwidth}
         \centering
         \includegraphics[width=\textwidth]{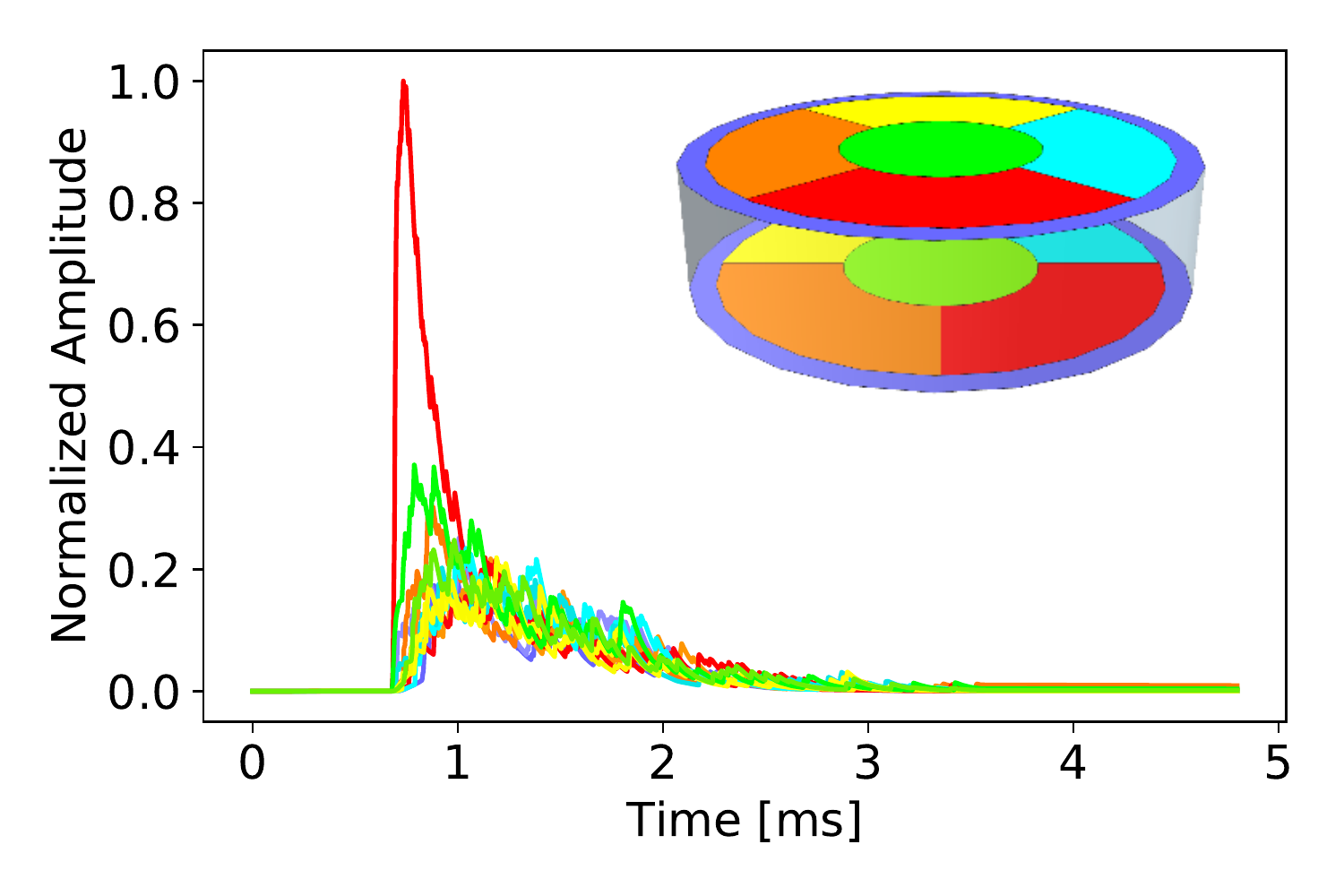}
         \caption{}
         \label{fig:DMCPulse}
     \end{subfigure}
     \hfill
     \begin{subfigure}[b]{0.49\textwidth}
         \centering
         \includegraphics[width=\textwidth]{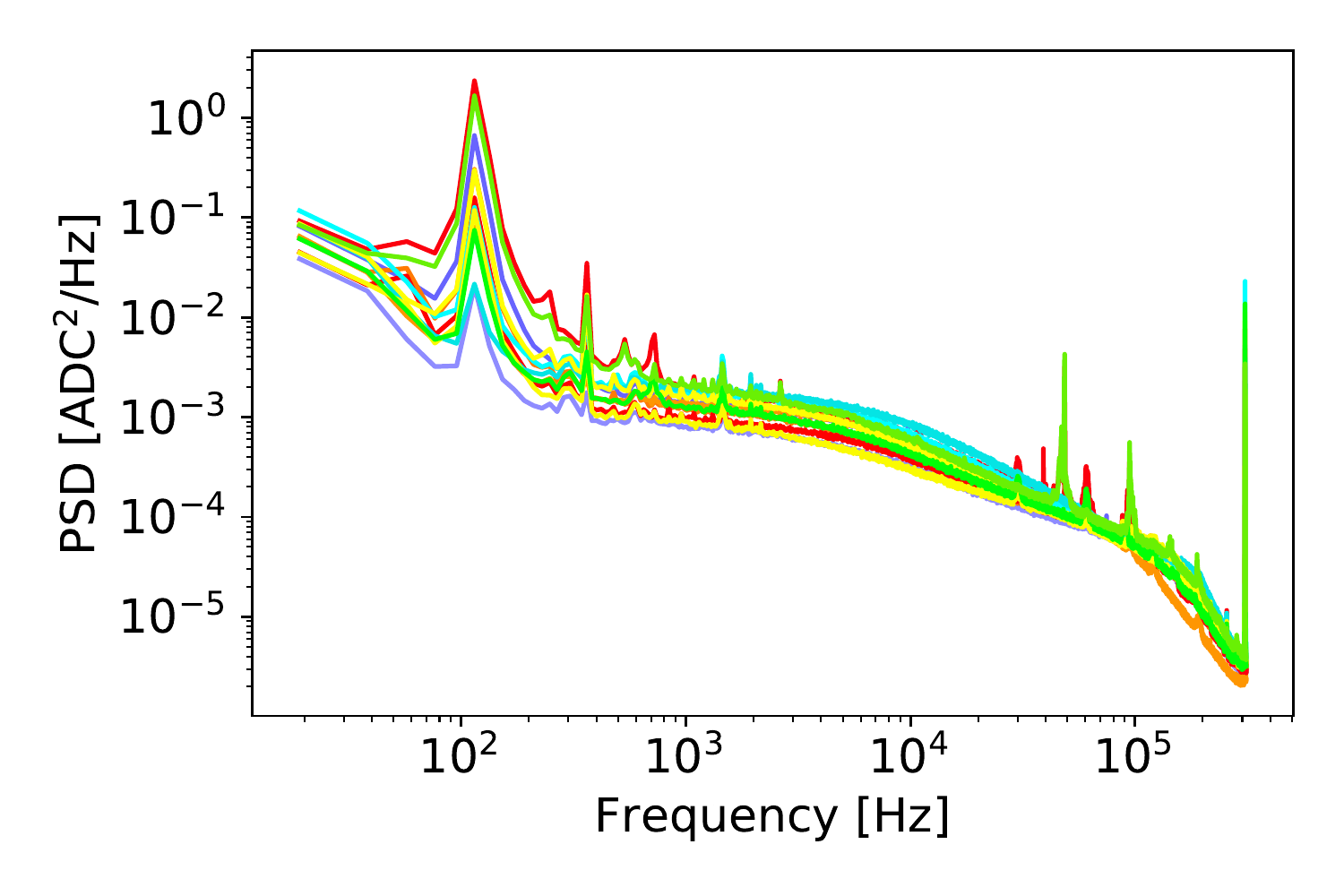}
         \caption{}
         \label{fig:DMCPSDs}
     \end{subfigure}
        \caption{Example phonon pulse (a) in the 12 phonon channels. The inserted sketch of an iZIP detector illustrates the color coding for each of the phonon channels. The noise power spectral densities (b) from randomly triggered noise traces from a prototype detector are shown for the same channels.}
        \label{fig:DMC}
\end{figure}

\begin{figure}[h]
     \centering
     \begin{subfigure}[b]{0.49\textwidth}
         \centering
         \includegraphics[width=\textwidth]{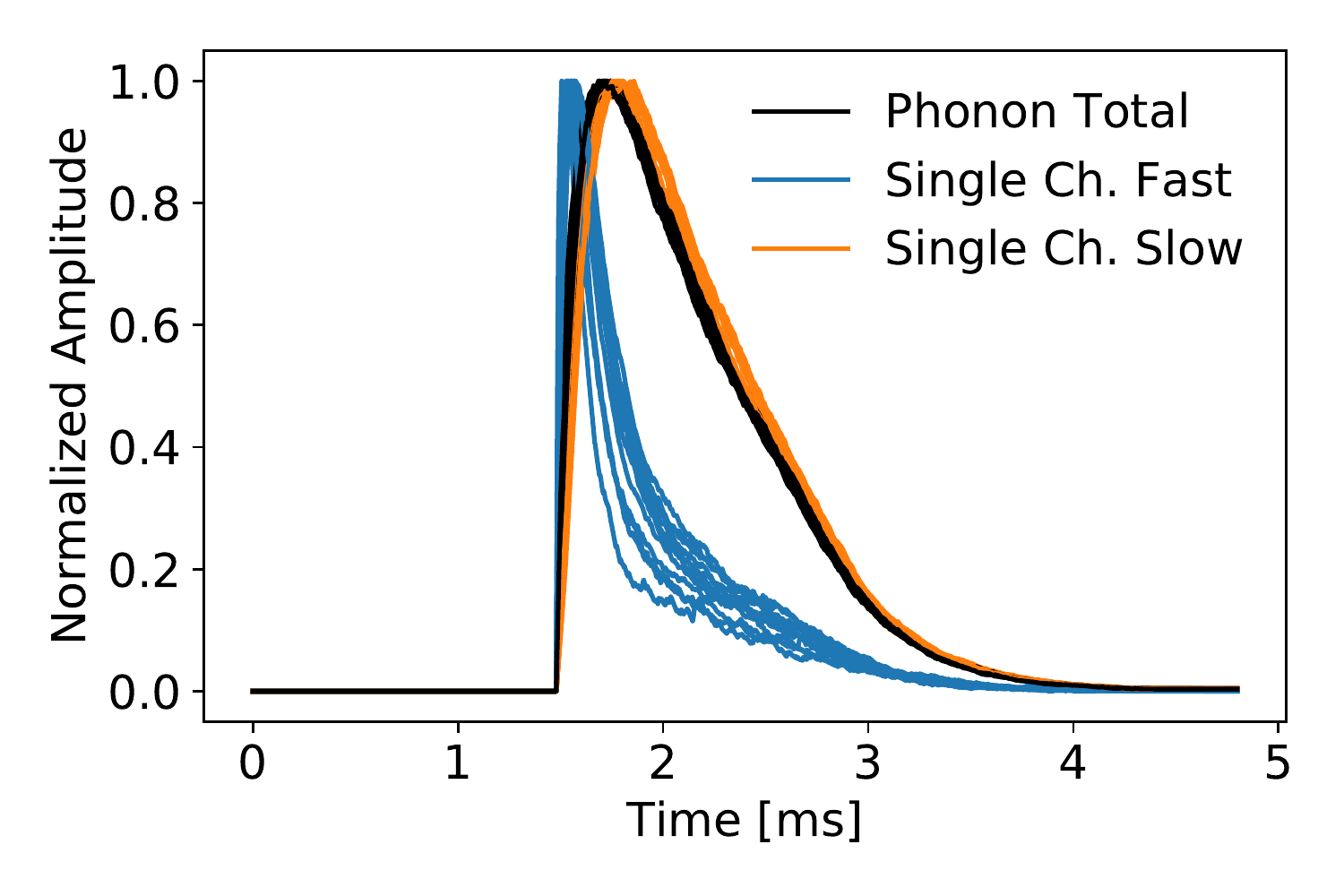}
         \caption{}
         \label{fig:DMCTemplates}
     \end{subfigure}
     \hfill
     \begin{subfigure}[b]{0.49\textwidth}
         \centering
         \includegraphics[width=\textwidth]{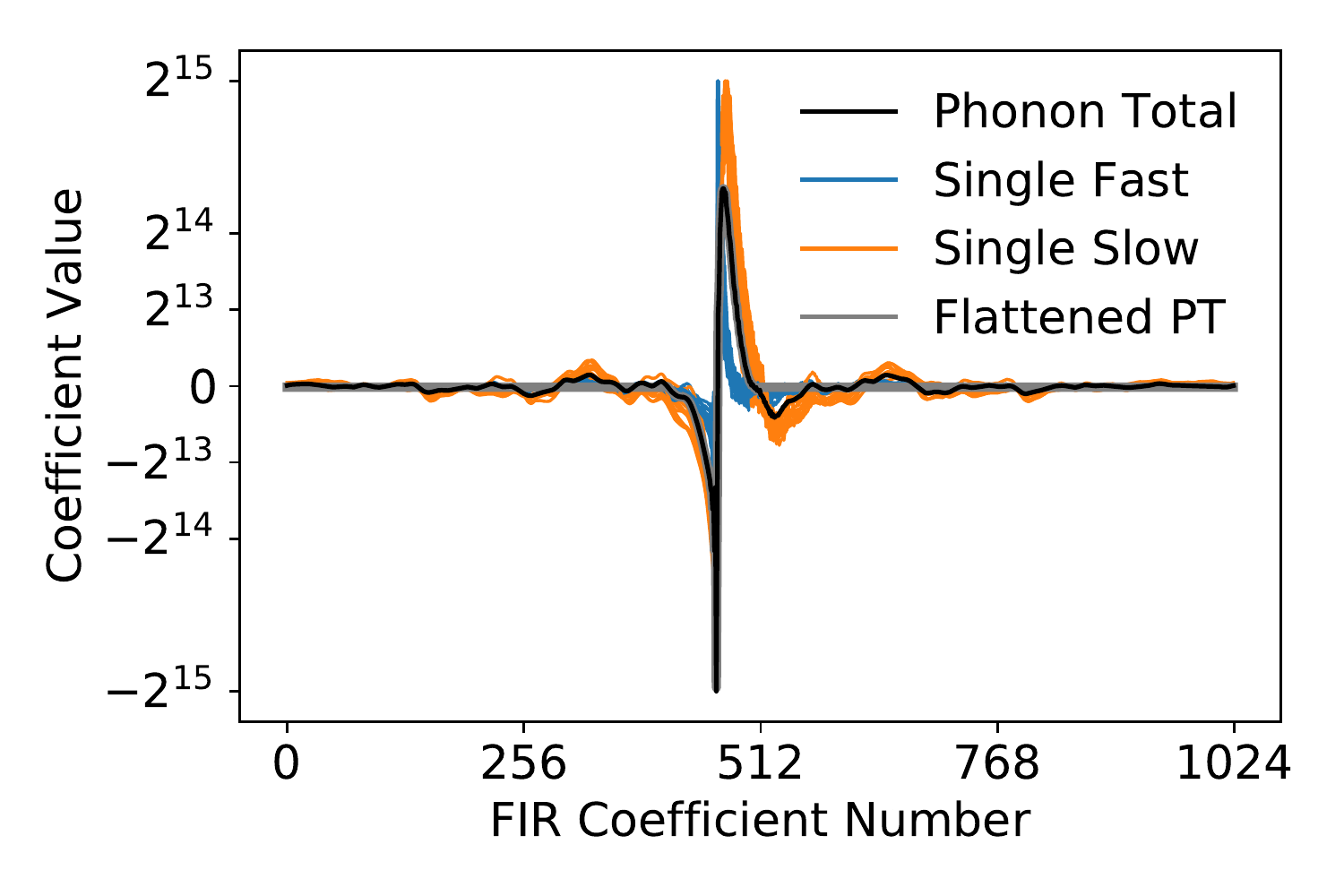}
         \caption{}
         \label{fig:FIRcoeffs}
     \end{subfigure}
        \caption{Phonon pulse templates (a) from fast (blue) and slow (orange) pulse components per individual channel as well as the total channel template (black). From each template, a corresponding set of optimal filter FIR coefficients (b) is derived. Additionally the coefficients for an optimal filter with flattened sidebands is shown (gray).}
        \label{fig:R85}
\end{figure}

The phonon pulses were taken from a \acrshort{DMC} sample containing 1000 simulated events of the interaction of a 10 GeV Weakly Interacting Massive Particle (\acrshort{WIMP}) with a SuperCDMS SNOLAB germanium crystal detector. One \acrshort{DMC} event contains the simulated signal pulses in the transition-edge sensors and as such is free of the electronic noise of the readout and data acquisition system. Due to the relatively high \acrshort{WIMP} mass the resulting pulses span a broad range of amplitudes. Figure~\ref{fig:DMC} (a) shows a \acrshort{DMC} pulse as read out in the 12 phonon channels. The pulse has fast and slowly rising and falling components which have different amplitudes depending on the proximity of the interaction to individual channels. The fast components stem from direct phonon hits and are only significant in channels with sensors close to the underlying recoil interaction in the detector. The slow components are created from additional ballistically scattering phonons \cite{watkins2023athermal} and have similar amplitudes in all channels. 
The shown pulse features a typical signal-like topology with a dominant fast rising pulse component in only one channel and smaller slow rising pulse components in the remaining channels. For each channel, template pulses are created from all \acrshort{DMC} events, as shown in figure~\ref{fig:R85} (a). The default templates are simply the averaged and normalized channel-specific pulses. The fast templates are created only from the channels with the highest amplitude per event and the slow templates from the remaining channels.

The noise traces were measured as randomly triggered traces from a surface run at a SLAC testing facility using a SuperCDMS SNOLAB \acrshort{iZIP} prototype germanium crystal detector. These noise traces provide a suitable environment to demonstrate the neural trigger performance in realistic noise conditions. Due to the absence of shielding, however, the traces are not free from pulses induced by cosmic radiation, which complicates the true pulse assignment and thus hinders both the \acrshort{NN} training and the performance evaluation. Each noise trace has a length of 32768 \acrshort{ADC} samples.
For this study an \acrshort{ADC} level preselection is employed to reject most of the noise traces containing large pulses from cosmic radiation. This is done by first removing the slope of the noise traces and then requiring the noise traces to have \acrshort{ADC} counts no greater than 3 times the noise traces' standard deviation. 
Small pulses cannot be removed to arbitrary precision by this method. A further selection requirement is derived based on the \acrshort{FIR} stage and is described in section~\ref{sec:FIRstage}. A total of 1175 selected slope-corrected traces are used to derive the neural trigger parameters as part of the training or validation sample.
The noise power spectral densities (\acrshort{PSD}s) across these selected traces are shown in figure \ref{fig:DMC} (b).

From the \acrshort{DMC} signal traces and the noise traces a set of separate signal+noise traces is created. For each noise trace two pulses are randomly chosen from the \acrshort{DMC} sample and are injected at a random position between the first and last quarter of the trace, but at a minimum distance of 1000 \acrshort{ADC} samples from each other to avoid overlap. The signal+noise, signal-only, as well as noise-only traces are used for different purposes during the derivation of the neural trigger parameters and the performance evaluation.

\subsection{FIR Stage}
\label{sec:FIRstage}

In the first step the pulse templates and \acrshort{PSD}s are used to derive the \acrshort{OF} coefficients for each trigger path. Afterwards the signal+noise traces are propagated through the trigger simulation up to and including the \acrshort{FIR} module in order to create the filtered traces, which serve as the input to the \acrshort{NN} module. 

\subsubsection{Filter Derivation}

To derive the \acrshort{OF} \acrshort{FIR} coefficients for the \acrshort{FIR} module, the pulse templates and \acrshort{PSD}s are propagated through the downsampling and linear combination modules in the trigger simulation. At the linear combination stage the resulting pulse templates and \acrshort{PSD}s are used to derive the \acrshort{OF} \acrshort{FIR} coefficients according to the procedure described in ref.~\cite{Wilson_2022}. For the second trigger path the \acrshort{FIR} coefficients in the sidebands are manually flattened resulting in a \acrshort{FOF}. The resulting \acrshort{FIR} coefficients are shown in figure~\ref{fig:R85} (b).

\subsubsection{FIR Output}

\begin{figure}[h]
    \centering
    \includegraphics[width=0.98\textwidth]{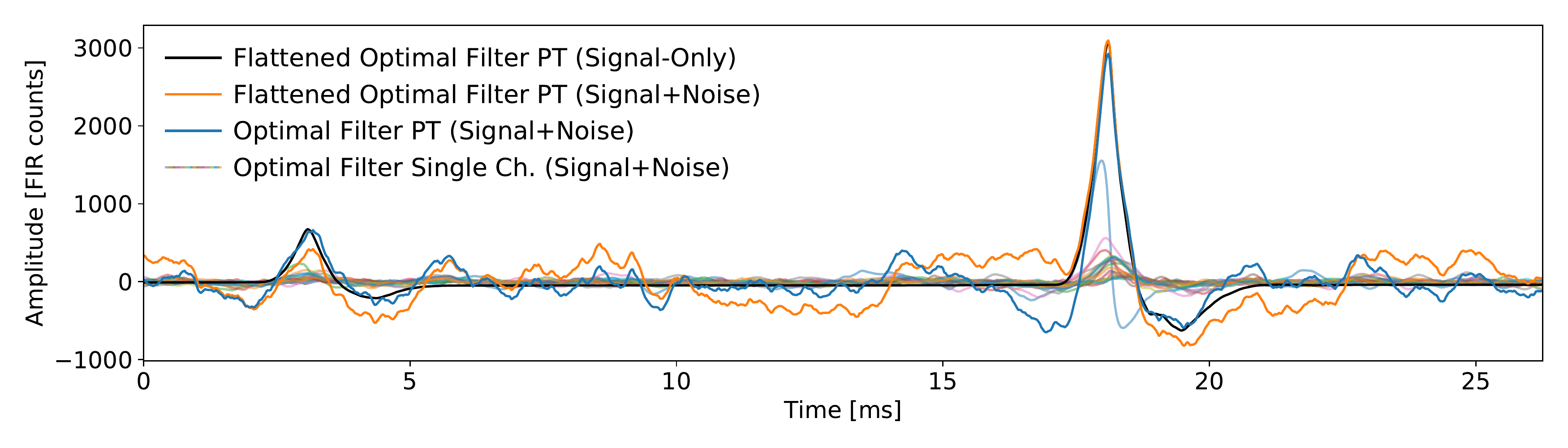}
    \caption{Outputs of the FIR module for the signal+noise and signal-only inputs. The signal+noise optimal filter (blue) and flattened optimal filter (orange) outputs from the phonon total (PT) channel serve as the inputs for the neural network training. The signal-only flattened optimal filter (FOF) output from the PT channel is the neural network training target. The various translucent traces are the signal+noise optimal filter outputs from the individual channels and are inputs to the neural network as well.}
    \label{fig:R85FIRtraces}
\end{figure}

The input traces, i.e., the signal+noise traces, the signal-only traces, as well as the noise-only traces are propagated through the downsampling, linear combination, and \acrshort{FIR} modules of the trigger simulation. Figure~\ref{fig:R85FIRtraces} shows the \acrshort{FIR} outputs for an input signal+noise trace, as well as the \acrshort{FOF} \acrshort{FIR} output for the signal-only component from the same trace. The \acrshort{FOF} trigger path output of the signal-only trace will be used as the target trace to which the \acrshort{NN} weights are fitted. The \acrshort{FOF} \acrshort{PT} provides a useful target trace because it is inherently similar to the \acrshort{OF} \acrshort{PT} output without the occurrence of oscillations leading to the echo trigger behavior described in ref.~\cite{Wilson_2022}. The noise-only \acrshort{OF} \acrshort{PT} output traces are used for a further selection criterion to remove noise traces containing residual pulses. The noise traces' maximum \acrshort{OF} \acrshort{PT} output value is required to be smaller than 5 times the \acrshort{OF} \acrshort{PT} baseline resolution.

\subsection{Neural Network}

In a training procedure the weights of the neurons are fitted to the \acrshort{FOF} \acrshort{PT} output of the signal-only traces. Once the weights have been determined, they are loaded in the \acrshort{NN} module, which finalizes the neural trigger setup. The combined filtered output trace is determined by propagating input traces up through the \acrshort{NN} module in the trigger simulation.

\subsubsection{Training}
\label{sec:training}

For the \acrshort{NN} training, the labeling of the input samples should be as pure as possible, i.e., ideally there should be no residual pulses within the noise traces.
Since not all residual pulses can be removed by the preselection criteria, the noise traces are inverted, i.e., multiplied by $-1$. Therefore any residual small pulses will become small dips, while the noise, which is assumed to be symmetric\footnote{The noise symmetry has been verified on a testbench setup, however potential asymmetric glitches might appear in future operations and should be monitored.}, will not be affected. The inverted input samples are split into training and validation samples with proportions 75:25. The \acrshort{NN} inputs are the \acrshort{FIR} outputs from the signal+noise training traces. They are normalized via division by a unit power of two which is closest to their standard deviation. 
The weights are fitted to the \acrshort{FOF} \acrshort{PT} output of the signal-only training traces using the maximum square loss function and the ADAM~\cite{https://doi.org/10.48550/arxiv.1412.6980} optimizer configured with an initial learning rate of $10^{-2}$. Once the validation loss reaches a plateau for at least two epochs the learning rate is decreased stepwise down to a learning rate of $10^{-5}$. After 339 training epochs the validation loss did not improve further and the \acrshort{NN} was considered fully trained. The final Keras model is converted to an HLS4ML model which is used to configure the \acrshort{NN} module. The outputs of the HLS4ML model and the Keras model agree with each other to within 0.5\%.

\subsubsection{Network Output}

\begin{figure}[htbp]
    \centering
    \includegraphics[width=0.98\textwidth]{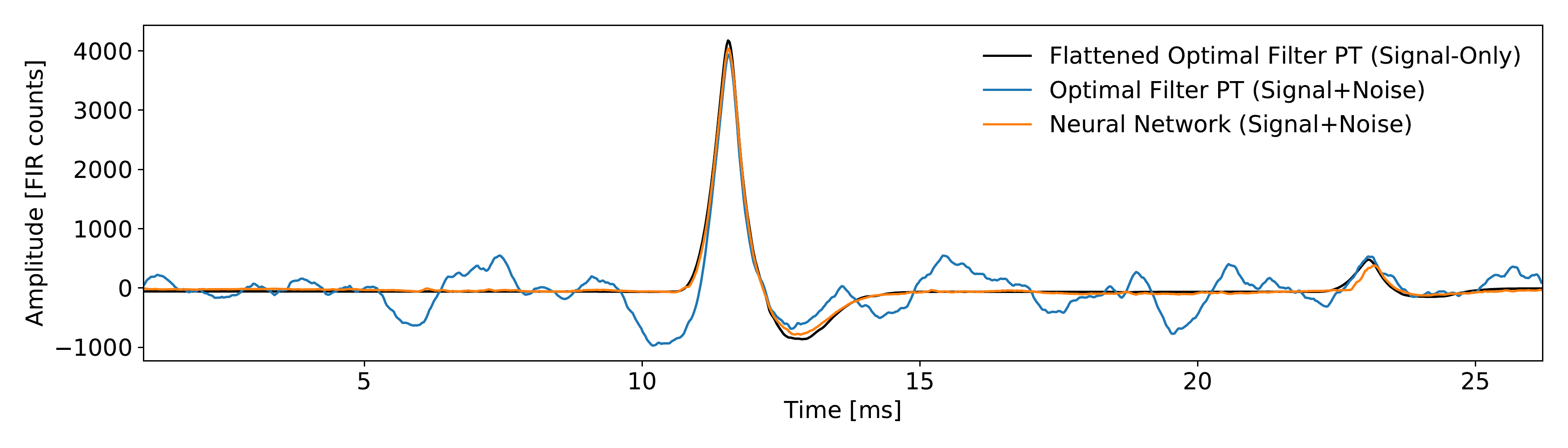}
    \caption{Neural network output (orange) compared to the optimal filter output (blue) for an example signal+noise phonon total (PT) trace, as well as the flattened optimal filter (FOF) output (black) for the corresponding signal-only PT trace.}
    \label{fig:R85_2GeV_Ge_NNout}
\end{figure}

To illustrate the \acrshort{NN} output, a signal+noise trace from a statistically independent test sample is sent through the data pipeline described above. Figure~\ref{fig:R85_2GeV_Ge_NNout} shows the \acrshort{NN} output compared to the \acrshort{OF} \acrshort{PT} output for an example signal+noise trace. The \acrshort{FOF} \acrshort{PT} output for the corresponding signal-only trace is also shown. The signal-to-noise ratio for the \acrshort{NN} appears significantly improved over the \acrshort{OF} \acrshort{PT} output. 
The deviation of the \acrshort{FOF} \acrshort{PT} and the \acrshort{OF} \acrshort{PT} in the pre pulse region is related to the shape of the \acrshort{FOF} \acrshort{PT} \acrshort{FIR} coefficients which are frozen when the peak tail of the \acrshort{OF} \acrshort{PT} \acrshort{FIR} coefficients crosses zero. 
The overall improvement is quantified in the following section~\ref{sec:performance}.

\section{Performance}
\label{sec:performance}
The performance of the neural trigger is quantified in terms of the trigger resolution, efficiency, turn-on point, and noise rate. This is done based on 1125 noise-only and signal+noise traces from a statistically independent test sample with the same preselection and inversion procedure applied as for the training and validation sample. The traces are sent through the data pipeline described in section~\ref{sec:inputdata} and compared to the output of the corresponding signal-only traces.

\subsection{Baseline Resolution}
The baseline resolution is determined by feeding noise-only traces through the data pipeline and measuring the standard deviation $\sigma$ of the distribution of sample values from the \acrshort{NN} output or \acrshort{OF} \acrshort{PT} output. The respective output sample distributions are shown in figure~\ref{fig:res} (a). While the \acrshort{OF} \acrshort{PT} output distribution resembles a Gaussian to good approximation, the \acrshort{NN} output is clearly non-Gaussian, asymmetric and features a sharp peak around zero. The \acrshort{NN} has a tendency to predict values close to zero due to the training procedure during which the \acrshort{NN} is presented with mostly noise samples values biasing it to predict zero as the target value. The asymmetric shape also stems from the \acrshort{NN} training procedure where the target trace sample distribution is also asymmetric. For comparability and to account for the shape differences, the baseline resolution is calculated from the arithmetical standard deviation rather than from a fit. It amounts to $152.94 \pm 0.10$ FIR counts for the \acrshort{OF} \acrshort{PT} output and $22.987 \pm 0.015$ FIR counts for the \acrshort{NN} output. The maximum \acrshort{OF} \acrshort{PT} and \acrshort{NN} output amplitudes among all noise-only traces correspond to a noise-free trigger threshold for this dataset at 796 and 573 FIR counts, respectively.

\subsection{Signal Resolution}
\label{sec:sigres}
\begin{figure}[h]
     \centering
     \begin{subfigure}[b]{0.49\textwidth}
         \centering
         \includegraphics[width=\textwidth]{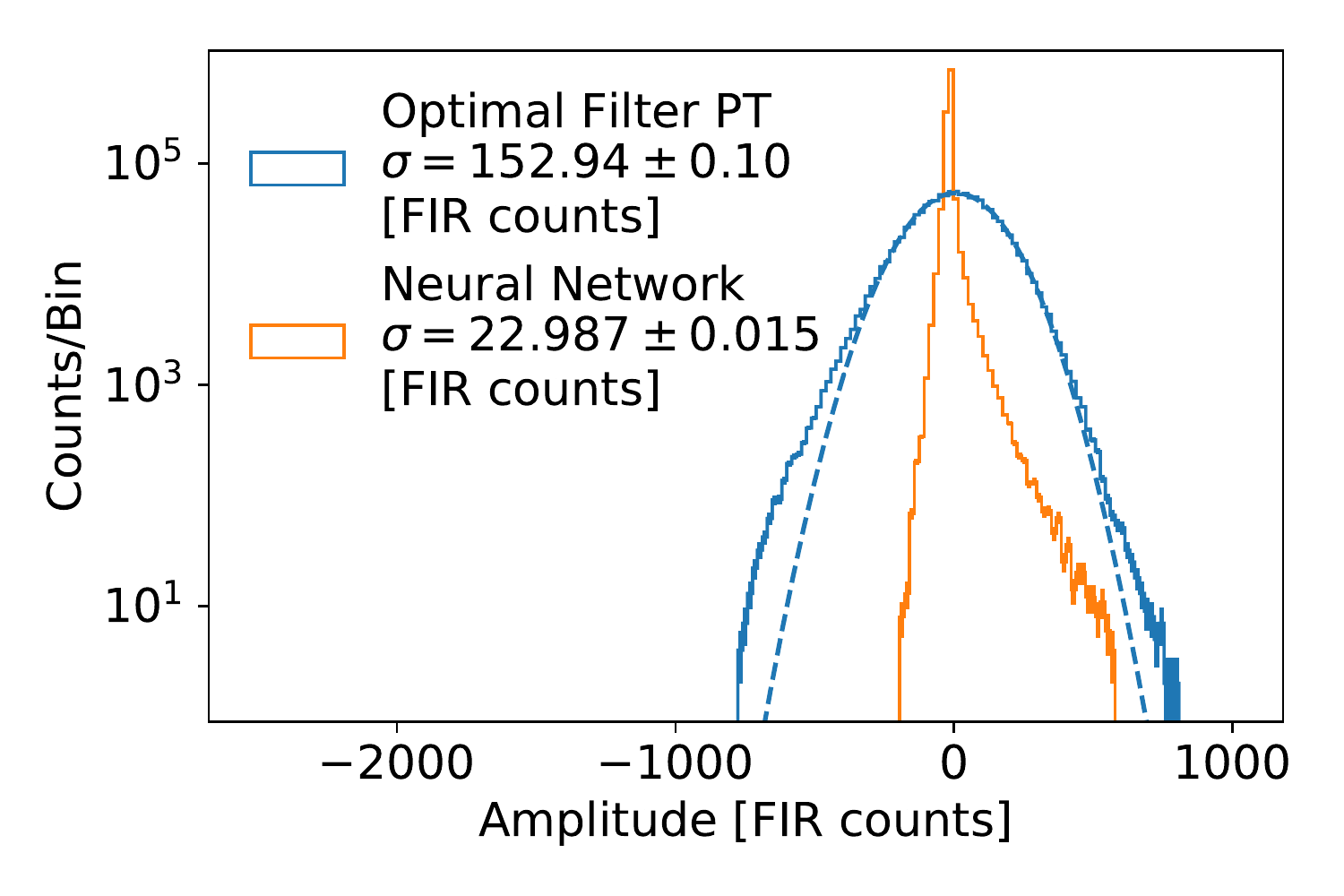}
         \caption{}
         \label{fig:baseres}
     \end{subfigure}
     \hfill
     \begin{subfigure}[b]{0.49\textwidth}
         \centering
        \includegraphics[width=\textwidth]{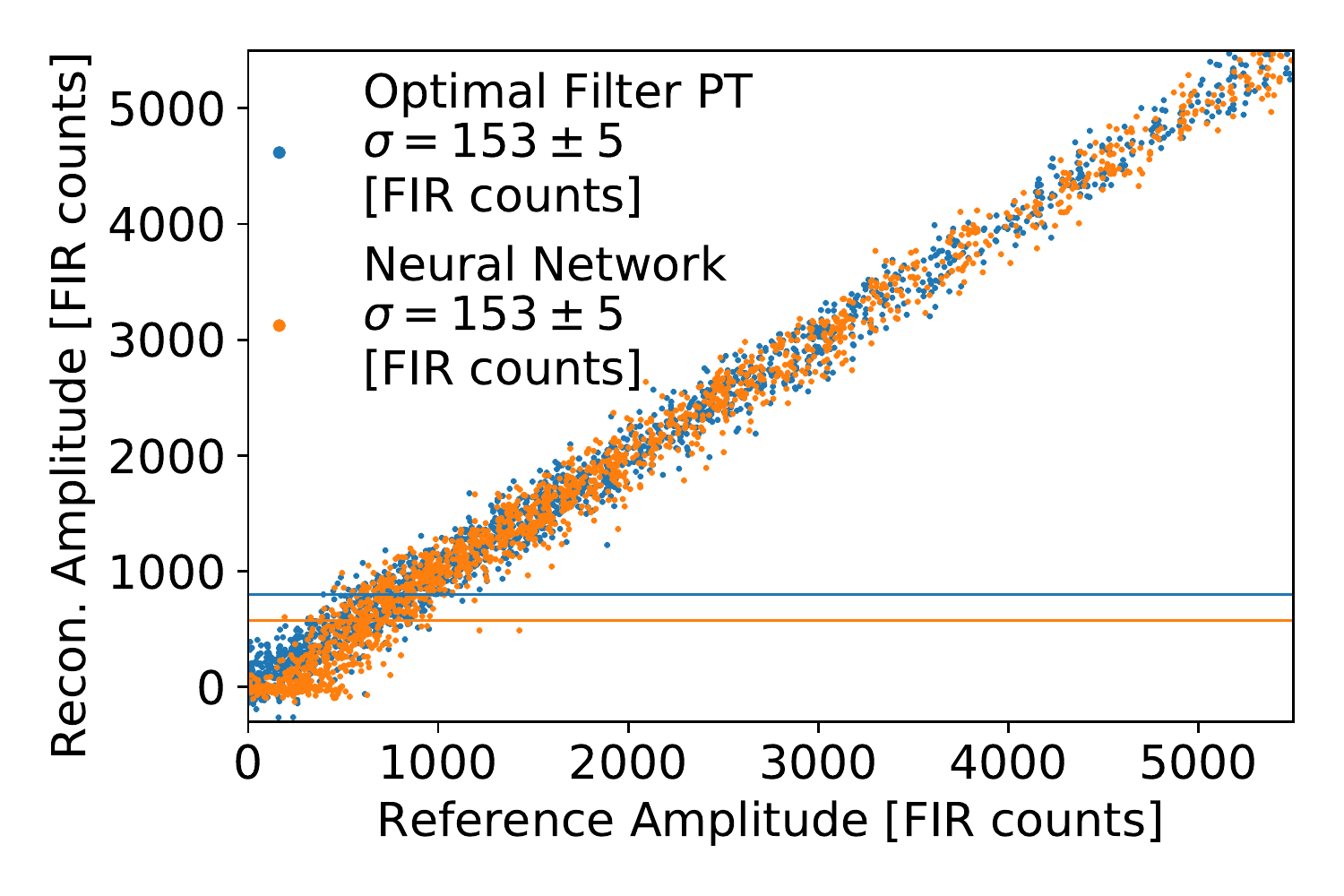}
         \caption{}
         \label{fig:sigres}
     \end{subfigure}
        \caption{Distribution of baseline amplitude values (a) across all noise samples which determine the baseline resolution for the neural network (orange) compared to the optimal filter (blue) acting on the phonon total (PT) trace. The blue dashed line shows a Gaussian fit to the optimal filter baseline distribution. The bin width corresponds to $\sim$18 FIR counts. To illustrate the signal resolution (b), the reconstructed signal peak amplitudes are shown as a function of the reference amplitudes. The horizontal solid lines mark the noise-free trigger thresholds.}
        \label{fig:res}
\end{figure}

The signal resolution is determined by comparing the reconstructed peak amplitudes from the signal+noise sample to the reference peak amplitudes from the respective signal-only sample. This way the reconstructed peak amplitudes will be affected by noise but the evaluation is free of noise triggers. Figure~\ref{fig:res} (b) shows the peak amplitudes of the \acrshort{OF} \acrshort{PT} and the \acrshort{NN} as a function of the reference amplitudes. The \acrshort{OF} \acrshort{PT} amplitudes are distributed linearly, whereas the \acrshort{NN} output amplitudes feature a flat region followed by a step at low amplitudes. This step, however, is far below the noise-free threshold (solid orange line) and has no practical implications on the \acrshort{NN} performance. Above this threshold, the \acrshort{NN} amplitudes are linearly distributed. The signal resolution is determined for amplitudes right above the noise-free trigger thresholds, i.e., up to 1500 FIR counts. It is calculated from the standard deviation of the distribution of the difference between reconstructed and reference amplitudes, amounting to $153 \pm 5$ for both the \acrshort{OF} \acrshort{PT} and the \acrshort{NN}. The similarity of the signal resolution suggests that the improved performance of the \acrshort{NN} is mainly driven by the improved baseline resolution and noise rejection.

\subsection{Trigger Efficiency and Thresholds}

To characterize the trigger efficiency, turn-on points, and noise rate quantitatively, the number of true and false triggers (noise triggers) need to be compared. This is done using a different method than in section~\ref{sec:sigres} to account for the actual peak finding behavior of the trigger system  and to be able to identify noise triggers. Accordingly, the reconstructed traces are propagated through the threshold logic, peak search, and trigger logic modules in the trigger simulation for varied threshold values and the identified trigger primitives are compared.
To calculate the number of noise triggers this is done using the noise-only \acrshort{OF} \acrshort{PT} and \acrshort{NN} output traces. Since the noise-only traces do not contain pulses, any trigger primitive associated to the \acrshort{OF} \acrshort{PT} or the \acrshort{NN} can be considered a noise trigger. 
To determine the number of true triggers, the signal+noise \acrshort{OF} \acrshort{PT} and \acrshort{NN} output traces are used in conjunction with the signal-only \acrshort{FOF} \acrshort{PT} output traces on a separate trigger path. The threshold value associated to the \acrshort{FOF} signal-only trigger path is kept constant with a minimal activation (deactivation) threshold of 2 (1) FIR counts. 
A true trigger is defined as the trigger primitive associated to the \acrshort{FOF} signal-only trigger path as it cannot be caused by an echo trigger. Therefore a true \acrshort{NN} or \acrshort{OF} \acrshort{PT} positive trigger requires a trigger primitive issued in the \acrshort{FOF} \acrshort{PT} signal-only trigger path and an exceeded threshold logic bit associated to the \acrshort{NN} or \acrshort{OF} \acrshort{PT}, respectively, during the trigger window. 

\begin{figure}[b]
     \centering
     \begin{subfigure}[b]{0.49\textwidth}
         \centering
         \includegraphics[width=\textwidth]{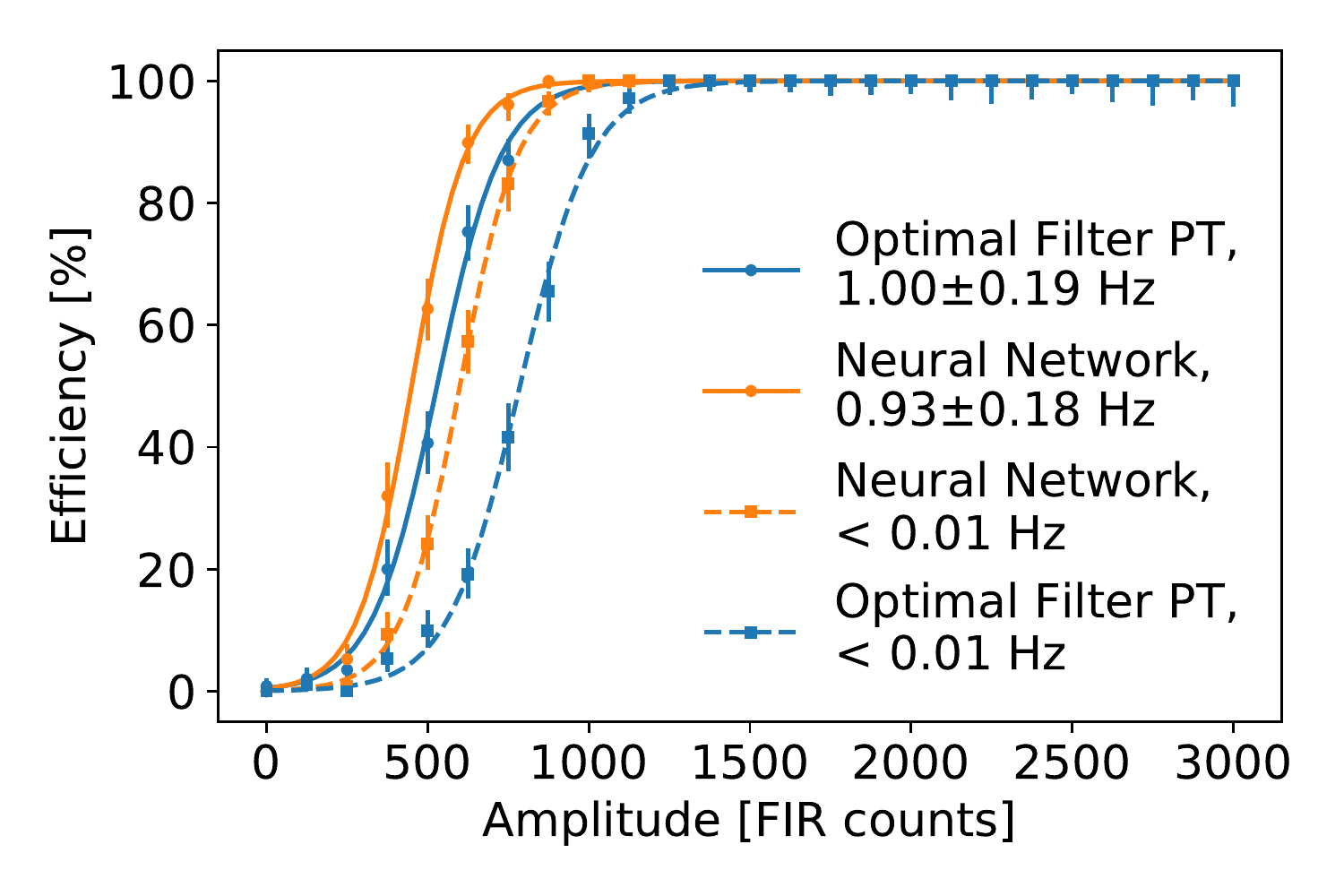}
         \caption{}
         \label{fig:efficiency}
     \end{subfigure}
     \hfill
     \begin{subfigure}[b]{0.49\textwidth}
         \centering
         \includegraphics[width=\textwidth]{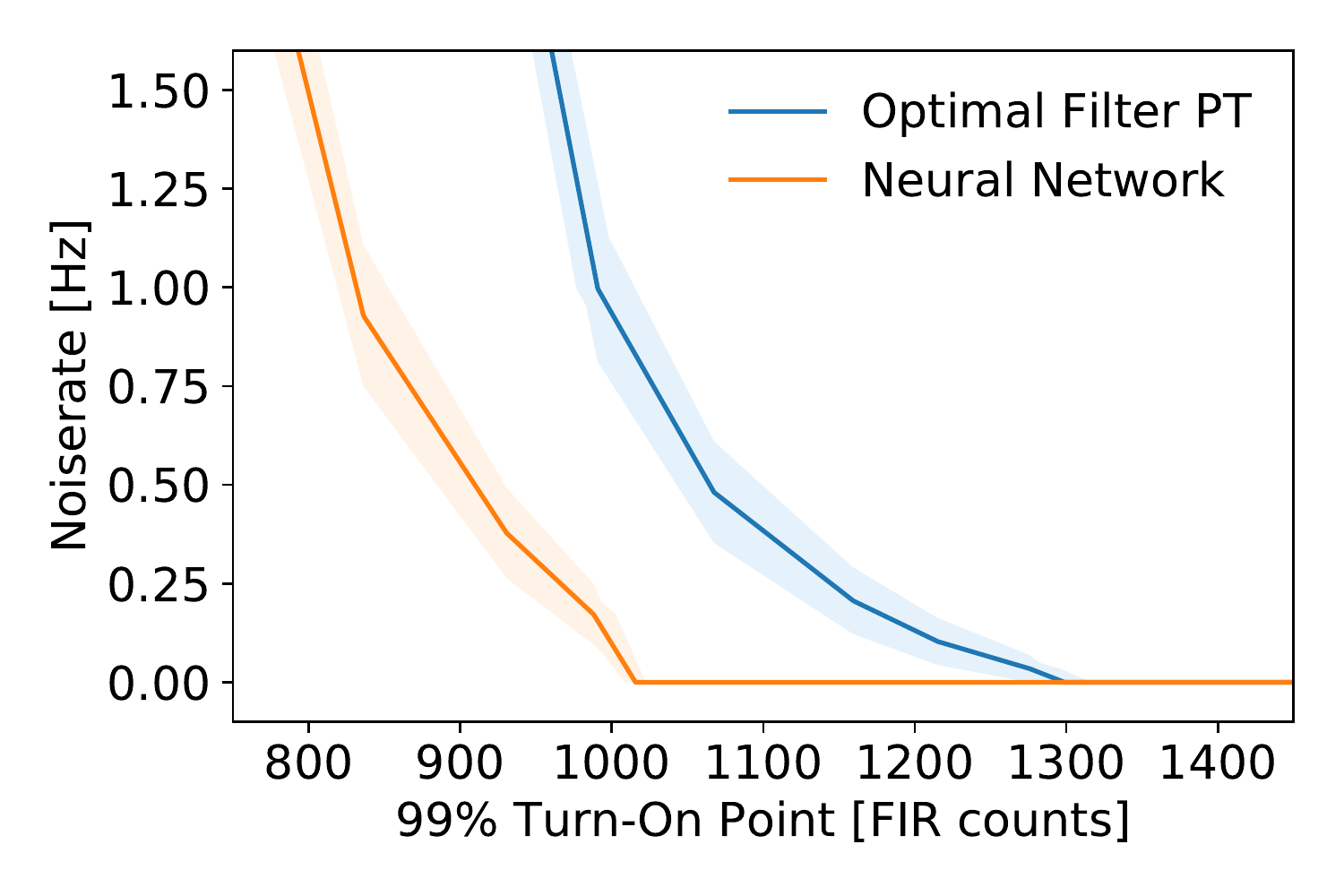}
         \caption{}
         \label{fig:noiseturnon}
     \end{subfigure}
        \caption{Trigger efficiency as a function of pulse amplitude (a) of the neural network (orange) compared to the optimal filter (blue) acting on the phonon total (PT) trace at two example thresholds which produce a similar noise rate noted in the legend. The noise rate as a function of the 99\% efficiency turn-on point at various thresholds is shown in (b), indicating an overall improved performance of the neural network compared to the optimal filter PT performance.}
        \label{fig:eff}
\end{figure}

From this classification, efficiency turn-on curves are determined and used to quantitatively characterize the performance.
An efficiency turn-on curve for a filter is created from the amplitude distribution of true positive triggers from the signal+noise traces divided by the amplitude distribution of all true triggers from the signal-only traces. The noise rate is determined by dividing the number of false positive triggers by the total trace length in the test sample. The efficiency turn-on curves are shown in figure~\ref{fig:eff} (a) for a few example thresholds where the \acrshort{OF} \acrshort{PT} and the \acrshort{NN} produce similar noise rates. Each curve is fitted with a sigmoid function.
The figure illustrates that, for a given noise rate, the \acrshort{NN} becomes fully efficient at lower pulse amplitudes compared to the \acrshort{OF} \acrshort{PT}. Additionally, the turn-on curve of the \acrshort{NN} is slightly steeper than the curve of the \acrshort{OF} \acrshort{PT}. Next, for each threshold data point the noise rate is determined and the point of 99\% efficiency is calculated from the sigmoid fit. Finally, for each threshold data point the noise rate is plotted against the point of 99\% efficiency, with the results shown in figure~\ref{fig:eff} (b). 
The plot shows that, for a given noise rate, the \acrshort{NN} is significantly more efficient than the \acrshort{OF} \acrshort{PT}. The 99\% efficiency turn-on point is at a significantly lower amplitude, and specifically is ${\sim}22$\% lower for a $<0.01$\,Hz noise rate working point. The uncertainties are taken from the fit errors of the sigmoid function and the statistical error on the number of noise triggers. Based on a coarse sensitivity estimation using a zero-background hypothesis, the improvement in the trigger threshold is expected to lower the smallest detectable \acrshort{WIMP} mass using germanium \acrshort{iZIP} detectors by ${\sim}12$\%. Additionally, given the exponential rise of the \acrshort{WIMP} signal spectrum towards lower energies, the sensitivity to previously observable low masses will improve notably. A detailed sensitivity analysis using underground data from all detector types remains to be performed in a future study.

\section{Conclusion}
\label{sec:conclusion}
The SuperCDMS level-1 trigger should reach the highest possible trigger efficiency at the lowest possible trigger threshold while maintaining an acceptable noise rate. In the default design of the level-1 trigger, the energy estimation of a putative pulse is made in separate trigger paths in a downsampling, linear combination, and \acrshort{FIR} filter module that is configured as an optimal filter. A combination of information from different trigger paths can happen only in the logic in terms of combinations of thresholds.
In this article a trigger upgrade towards a neural trigger is proposed in which the \acrshort{FIR} filtered traces from different trigger paths are combined using an Long Short-Term Memory neural network, resulting in an improved baseline resolution and signal efficiency and thus a lower possible trigger threshold at a given noise rate. This upgrade is examined using a software-based trigger simulation where the neural network is emulated using an HLS4ML model. The trigger paths of the neural trigger are configured with different optimal filters running on the total as well as individual phonon channels. The optimal filter configuration uses pulse templates describing both fast and slow rising pulse components in each channel, which allows the neural network to model the pulse topology in higher detail than in the default optimal filter. This improves the discrimination between signal- and noise-like features during the pulse amplitude estimation.
The neural network is trained using simulated pulses injected into randomly triggered noise traces measured with a prototype \acrshort{iZIP} detector. Noise traces containing significant pulses not stemming from the simulation are removed in a cut-based selection procedure. By inverting the remaining noise traces, any potential residual pulses are eliminated.
Using a trained and configured instance of the neural trigger in the trigger simulation, the baseline resolution and efficiency of the neural trigger is significantly improved allowing for a ${\sim}22$\% lower threshold than the optimal filter trigger at a $< 0.01$\,Hz noise rate.
These findings motivate the validation of the results on a broader basis, i.e. not only using germanium \acrshort{iZIP} detectors but also \acrshort{HV} and silicon detectors which have a considerably lower threshold. The studies should also be repeated with data from an underground test facility and using real detector pulses. Overall, the results encourage the hardware implementation of the neural network trigger module between the \acrshort{FIR} module and the threshold module in the SuperCDMS trigger \acrshort{FPGA}.

\acknowledgments

We would like to thank David Toback, Michael Kelsey, Rik Bhattacharyya and Joshua Winchell for their contributions to the DMC and trigger simulations. We are grateful to Francisco Ponce for the noise data taking. We thank Stella Wermuth and Matthew Wilson for comments on a draft version of this manuscript. We gratefully acknowledge support from the Deutsche Forschungsgemeinschaft (DFG, German Research Foundation) under the Emmy Noether Grant No. 420484612 and from the U.S.\ Department of Energy Office of High Energy Physics under the award number DE-SC0007861.

\bibliographystyle{iopart-num}
\bibliography{references}{}

\providecommand{\newblock}{}
\begin{thebibliography}{10}
\expandafter\ifx\csname url\endcsname\relax
  \def\url#1{{\tt #1}}\fi
\expandafter\ifx\csname urlprefix\endcsname\relax\def\urlprefix{URL }\fi
\providecommand{\eprint}[2][]{\url{#2}}

\bibitem{Agnese_2017}
Agnese R {\em et~al.\/} (SuperCDMS) Projected sensitivity of the {SuperCDMS}
  {SNOLAB} experiment 2017 {\em Physical Review D\/} {\bf 95} ISSN 2470-0029
  \urlprefix\url{http://dx.doi.org/10.1103/PhysRevD.95.082002}

\bibitem{Neganov:1985khw}
Neganov B~S and Trofimov V~N {Colorimetric method measuring ionizing radiation}
  1985 {\em Otkryt. Izobret.\/} {\bf 146} 215

\bibitem{Luke1988VoltageassistedCI}
Luke P~N Voltage‐assisted calorimetric ionization detector 1988 {\em Journal
  of Applied Physics\/} {\bf 64} 6858--6860

\bibitem{Wilson_2022}
Wilson J, {Meyer zu Theenhausen} H, von Krosigk B, Azadbakht E, Bunker R, Hall
  J, Hansen S, Hines B, Loer B, Olsen J, Oser S, Partridge R, Pyle M, Sander J,
  Serfass B, Toback D, Watkins S and Zhao X The level-1 trigger for the
  {SuperCDMS} experiment at {SNOLAB} 2022 {\em Journal of Instrumentation\/}
  {\bf 17} P07010
  \urlprefix\url{https://doi.org/10.1088/1748-0221/17/07/p07010}

\bibitem{MANCUSO2019492}
Mancuso M, Bento A, Iachellini N~F, Hauff D, Petricca F, Pr{\"o}bst F, Rothe J
  and Strauss R A method to define the energy threshold depending on noise
  level for rare event searches 2019 {\em Nuclear Instruments and Methods in
  Physics Research Section A: Accelerators, Spectrometers, Detectors and
  Associated Equipment\/} {\bf 940} 492--496 ISSN 0168-9002
  \urlprefix\url{https://www.sciencedirect.com/science/article/pii/S0168900219308708}

\bibitem{OF_gatti}
Gatti E and Manfredi P~F Processing the signals from solid-state detectors in
  elementary-particle physics 1986 {\em La Rivista del Nuovo Cimento
  (1978-1999)\/} {\bf 9} 1--146

\bibitem{hochreiter1997long}
Hochreiter S and Schmidhuber J Long short-term memory 1997 {\em Neural
  computation\/} {\bf 9} 1735--1780

\bibitem{agarap2018deep}
Agarap A~F Deep learning using rectified linear units (relu) 2018 {\em arXiv
  preprint arXiv:1803.08375\/}

\bibitem{chollet2015keras}
Chollet F {\em et~al.\/} 2015 Keras
  \urlprefix\url{https://github.com/fchollet/keras}

\bibitem{Duarte_2018}
Duarte J, Han S, Harris P, Jindariani S, Kreinar E, Kreis B, Ngadiuba J,
  Pierini M, Rivera R, Tran N and Wu Z Fast inference of deep neural networks
  in {FPGAs} for particle physics 2018 {\em Journal of Instrumentation\/} {\bf
  13} P07027--P07027
  \urlprefix\url{https://doi.org/10.1088%2F1748-0221%2F13%2F07%2Fp07027}

\bibitem{165599}
Sontag E Feedback stabilization using two-hidden-layer nets 1992 {\em IEEE
  Transactions on Neural Networks\/} {\bf 3} 981--990

\bibitem{https://doi.org/10.48550/arxiv.2006.10159}
Coelho C~N, Kuusela A, Li S, Zhuang H, Aarrestad T, Loncar V, Ngadiuba J,
  Pierini M, Pol A~A and Summers S Automatic heterogeneous quantization of deep
  neural networks for low-latency inference on the edge for particle detectors
  2020  \urlprefix\url{https://arxiv.org/abs/2006.10159}

\bibitem{watkins2023athermal}
Watkins S~L 2023 Athermal phonon sensors in searches for light dark matter
  (\textit{Preprint} \eprint{2301.08699})

\bibitem{https://doi.org/10.48550/arxiv.1412.6980}
Kingma D~P and Ba J 2014 Adam: A method for stochastic optimization
  \urlprefix\url{https://arxiv.org/abs/1412.6980}

\end{thebibliography}

\end{document}